\documentclass[twocolumn,aps,prc,showpacs,superscriptaddress,amsmath,amssymb,floatfix]{revtex4}

\usepackage{graphicx}
\usepackage{tabularx}
\usepackage{blopeps}
\blopps{1.5mm}
\bloplw{0.15mm}

\newcommand{\GeV}{\,\mathrm{GeV}}
\newcommand{\GeVpc}{\,\mathrm{GeV/c}}

\begin{document}
\title{High Transverse Momentum Hadron Spectra at 
$\sqrt{s_{{}_{NN}}}=17.3\GeV$, in Pb+Pb and p+p 
Collisions}





\affiliation{NIKHEF, Amsterdam, Netherlands.}
\affiliation{Department of Physics, University of Athens, Athens, Greece.}
\affiliation{Comenius University, Bratislava, Slovakia.}
\affiliation{KFKI Research Institute for Particle and Nuclear Physics, Budapest, Hungary.}
\affiliation{MIT, Cambridge, USA.}
\affiliation{Henryk Niewodniczanski Institute of Nuclear Physics, Polish Academy of Sciences, Cracow, Poland.}
\affiliation{Gesellschaft f\"{u}r Schwerionenforschung (GSI), Darmstadt, Germany.}
\affiliation{Joint Institute for Nuclear Research, Dubna, Russia.}
\affiliation{Fachbereich Physik der Universit\"{a}t, Frankfurt, Germany.}
\affiliation{CERN, Geneva, Switzerland.}
\affiliation{Institute of Physics \'Swi\c{e}tokrzyska Academy, Kielce, Poland.}
\affiliation{Fachbereich Physik der Universit\"{a}t, Marburg, Germany.}
\affiliation{Max-Planck-Institut f\"{u}r Physik, Munich, Germany.}
\affiliation{Charles University, Faculty of Mathematics and Physics, Institute of Particle and Nuclear Physics, Prague, Czech Republic.}
\affiliation{Department of Physics, Pusan National University, Pusan, Republic of Korea.}
\affiliation{Nuclear Physics Laboratory, University of Washington, Seattle, WA, USA.}
\affiliation{Atomic Physics Department, Sofia University St. Kliment Ohridski, Sofia, Bulgaria.} 
\affiliation{Institute for Nuclear Research and Nuclear Energy, Sofia, Bulgaria.} 
\affiliation{Department of Chemistry, Stony Brook Univ. (SUNYSB), Stony Brook, USA.}
\affiliation{Institute for Nuclear Studies, Warsaw, Poland.}
\affiliation{Institute for Experimental Physics, University of Warsaw, Warsaw, Poland.}
\affiliation{Faculty of Physics, Warsaw University of Technology, Warsaw, Poland.}
\affiliation{Rudjer Boskovic Institute, Zagreb, Croatia.}


\author{C.~Alt}\affiliation{Fachbereich Physik der Universit\"{a}t, Frankfurt, Germany.}
\author{T.~Anticic}\affiliation{Rudjer Boskovic Institute, Zagreb, Croatia.}
\author{B.~Baatar}\affiliation{Joint Institute for Nuclear Research, Dubna, Russia.}
\author{D.~Barna}\affiliation{KFKI Research Institute for Particle and Nuclear Physics, Budapest, Hungary.}
\author{J.~Bartke}\affiliation{Henryk Niewodniczanski Institute of Nuclear Physics, Polish Academy of Sciences, Cracow, Poland.}
\author{L.~Betev}\affiliation{CERN, Geneva, Switzerland.}
\author{H.~Bia{\l}\-kowska}\affiliation{Institute for Nuclear Studies, Warsaw, Poland.}
\author{C.~Blume}\affiliation{Fachbereich Physik der Universit\"{a}t, Frankfurt, Germany.}
\author{B.~Boimska}\affiliation{Institute for Nuclear Studies, Warsaw, Poland.}
\author{M.~Botje}\affiliation{NIKHEF, Amsterdam, Netherlands.}
\author{J.~Bracinik}\affiliation{Comenius University, Bratislava, Slovakia.}
\author{R.~Bramm}\affiliation{Fachbereich Physik der Universit\"{a}t, Frankfurt, Germany.}
\author{P.~Bun\v{c}i\'{c}}\affiliation{CERN, Geneva, Switzerland.}
\author{V.~Cerny}\affiliation{Comenius University, Bratislava, Slovakia.}
\author{P.~Christakoglou}\affiliation{Department of Physics, University of Athens, Athens, Greece.}
\author{P.~Chung}\affiliation{Department of Chemistry, Stony Brook Univ. (SUNYSB), Stony Brook, USA.}
\author{O.~Chvala}\affiliation{Charles University, Faculty of Mathematics and Physics, Institute of Particle and Nuclear Physics, Prague, Czech Republic.}
\author{J.G.~Cramer}\affiliation{Nuclear Physics Laboratory, University of Washington, Seattle, WA, USA.}
\author{P.~Csat\'{o}}\affiliation{KFKI Research Institute for Particle and Nuclear Physics, Budapest, Hungary.}
\author{P.~Dinkelaker}\affiliation{Fachbereich Physik der Universit\"{a}t, Frankfurt, Germany.}
\author{V.~Eckardt}\affiliation{Max-Planck-Institut f\"{u}r Physik, Munich, Germany.}
\author{D.~Flierl}\affiliation{Fachbereich Physik der Universit\"{a}t, Frankfurt, Germany.}
\author{Z.~Fodor}\affiliation{KFKI Research Institute for Particle and Nuclear Physics, Budapest, Hungary.}
\author{P.~Foka}\affiliation{Gesellschaft f\"{u}r Schwerionenforschung (GSI), Darmstadt, Germany.}
\author{V.~Friese}\affiliation{Gesellschaft f\"{u}r Schwerionenforschung (GSI), Darmstadt, Germany.}
\author{J.~G\'{a}l}\affiliation{KFKI Research Institute for Particle and Nuclear Physics, Budapest, Hungary.}
\author{M.~Ga\'zdzicki}\affiliation{Fachbereich Physik der Universit\"{a}t, Frankfurt, Germany.}\affiliation{Institute of Physics \'Swi\c{e}tokrzyska Academy, Kielce, Poland.}
\author{V.~Genchev}\affiliation{Institute for Nuclear Research and Nuclear Energy, Sofia, Bulgaria.} 
\author{G.~Georgopoulos}\affiliation{Department of Physics, University of Athens, Athens, Greece.}
\author{E.~G{\l}adysz}\affiliation{Henryk Niewodniczanski Institute of Nuclear Physics, Polish Academy of Sciences, Cracow, Poland.}
\author{K.~Grebieszkow}\affiliation{Faculty of Physics, Warsaw University of Technology, Warsaw, Poland.}
\author{S.~Hegyi}\affiliation{KFKI Research Institute for Particle and Nuclear Physics, Budapest, Hungary.}
\author{C.~H\"{o}hne}\affiliation{Gesellschaft f\"{u}r Schwerionenforschung (GSI), Darmstadt, Germany.}
\author{K.~Kadija}\affiliation{Rudjer Boskovic Institute, Zagreb, Croatia.}
\author{A.~Karev}\affiliation{Max-Planck-Institut f\"{u}r Physik, Munich, Germany.}
\author{D.~Kikola}\affiliation{Faculty of Physics, Warsaw University of Technology, Warsaw, Poland.}
\author{M.~Kliemant}\affiliation{Fachbereich Physik der Universit\"{a}t, Frankfurt, Germany.}
\author{S.~Kniege}\affiliation{Fachbereich Physik der Universit\"{a}t, Frankfurt, Germany.}
\author{V.I.~Kolesnikov}\affiliation{Joint Institute for Nuclear Research, Dubna, Russia.}
\author{E.~Kornas}\affiliation{Henryk Niewodniczanski Institute of Nuclear Physics, Polish Academy of Sciences, Cracow, Poland.}
\author{R.~Korus}\affiliation{Institute of Physics \'Swi\c{e}tokrzyska Academy, Kielce, Poland.}
\author{M.~Kowalski}\affiliation{Henryk Niewodniczanski Institute of Nuclear Physics, Polish Academy of Sciences, Cracow, Poland.}
\author{I.~Kraus}\affiliation{Gesellschaft f\"{u}r Schwerionenforschung 
(GSI), Darmstadt, Germany.}
\author{M.~Kreps}\affiliation{Comenius University, Bratislava, Slovakia.}
\author{A.~Laszlo}~\email[Corresponding author. E-mail address: ]{laszloa@rmki.kfki.hu}\affiliation{KFKI Research Institute for Particle and Nuclear Physics, Budapest, Hungary.}
\author{R.~Lacey}\affiliation{Department of Chemistry, Stony Brook Univ. (SUNYSB), Stony Brook, USA.}
\author{M.~van~Leeuwen}\affiliation{NIKHEF, Amsterdam, Netherlands.}
\author{P.~L\'{e}vai}\affiliation{KFKI Research Institute for Particle and Nuclear Physics, Budapest, Hungary.}
\author{L.~Litov}\affiliation{Atomic Physics Department, Sofia University St. Kliment Ohridski, Sofia, Bulgaria.} 
\author{B.~Lungwitz}\affiliation{Fachbereich Physik der Universit\"{a}t, Frankfurt, Germany.}
\author{M.~Makariev}\affiliation{Atomic Physics Department, Sofia University St. Kliment Ohridski, Sofia, Bulgaria.} 
\author{A.I.~Malakhov}\affiliation{Joint Institute for Nuclear Research, Dubna, Russia.}
\author{M.~Mateev}\affiliation{Atomic Physics Department, Sofia University St. Kliment Ohridski, Sofia, Bulgaria.} 
\author{G.L.~Melkumov}\affiliation{Joint Institute for Nuclear Research, Dubna, Russia.}
\author{A.~Mischke}\affiliation{NIKHEF, Amsterdam, Netherlands.}
\author{M.~Mitrovski}\affiliation{Fachbereich Physik der Universit\"{a}t, Frankfurt, Germany.}
\author{J.~Moln\'{a}r}\affiliation{KFKI Research Institute for Particle and Nuclear Physics, Budapest, Hungary.}
\author{St.~Mr\'owczy\'nski}\affiliation{Institute of Physics \'Swi\c{e}tokrzyska Academy, Kielce, Poland.}
\author{V.~Nicolic}\affiliation{Rudjer Boskovic Institute, Zagreb, Croatia.}
\author{G.~P\'{a}lla}\affiliation{KFKI Research Institute for Particle and Nuclear Physics, Budapest, Hungary.}
\author{A.D.~Panagiotou}\affiliation{Department of Physics, University of Athens, Athens, Greece.}
\author{D.~Panayotov}\affiliation{Atomic Physics Department, Sofia University St. Kliment Ohridski, Sofia, Bulgaria.} 
\author{A.~Petridis}~\email[Deceased.]{}\affiliation{Department of Physics, University of Athens, Athens, Greece.}
\author{W.~Peryt}\affiliation{Faculty of Physics, Warsaw University of Technology, Warsaw, Poland.}
\author{M.~Pikna}\affiliation{Comenius University, Bratislava, Slovakia.}
\author{J.~Pluta}\affiliation{Faculty of Physics, Warsaw University of Technology, Warsaw, Poland.}
\author{D.~Prindle}\affiliation{Nuclear Physics Laboratory, University of Washington, Seattle, WA, USA.}
\author{F.~P\"{u}hlhofer}\affiliation{Fachbereich Physik der Universit\"{a}t, Marburg, Germany.}
\author{R.~Renfordt}\affiliation{Fachbereich Physik der Universit\"{a}t, Frankfurt, Germany.}
\author{C.~Roland}\affiliation{MIT, Cambridge, USA.}
\author{G.~Roland}\affiliation{MIT, Cambridge, USA.}
\author{M. Rybczy\'nski}\affiliation{Institute of Physics \'Swi\c{e}tokrzyska Academy, Kielce, Poland.}
\author{A.~Rybicki}\affiliation{Henryk Niewodniczanski Institute of Nuclear Physics, Polish Academy of Sciences, Cracow, Poland.}
\author{A.~Sandoval}\affiliation{Gesellschaft f\"{u}r Schwerionenforschung (GSI), Darmstadt, Germany.}
\author{N.~Schmitz}\affiliation{Max-Planck-Institut f\"{u}r Physik, Munich, Germany.}
\author{T.~Schuster}\affiliation{Fachbereich Physik der Universit\"{a}t, Frankfurt, Germany.}
\author{P.~Seyboth}\affiliation{Max-Planck-Institut f\"{u}r Physik, Munich, Germany.}
\author{F.~Sikl\'{e}r}\affiliation{KFKI Research Institute for Particle and Nuclear Physics, Budapest, Hungary.}
\author{B.~Sitar}\affiliation{Comenius University, Bratislava, Slovakia.}
\author{E.~Skrzypczak}\affiliation{Institute for Experimental Physics, University of Warsaw, Warsaw, Poland.}
\author{M.~Slodkowski}\affiliation{Faculty of Physics, Warsaw University of Technology, Warsaw, Poland.}
\author{G.~Stefanek}\affiliation{Institute of Physics \'Swi\c{e}tokrzyska Academy, Kielce, Poland.}
\author{R.~Stock}\affiliation{Fachbereich Physik der Universit\"{a}t, Frankfurt, Germany.}
\author{C.~Strabel}\affiliation{Fachbereich Physik der Universit\"{a}t, Frankfurt, Germany.}
\author{H.~Str\"{o}bele}\affiliation{Fachbereich Physik der Universit\"{a}t, Frankfurt, Germany.}
\author{T.~Susa}\affiliation{Rudjer Boskovic Institute, Zagreb, Croatia.}
\author{I.~Szentp\'{e}tery}\affiliation{KFKI Research Institute for Particle and Nuclear Physics, Budapest, Hungary.}
\author{J.~Sziklai}\affiliation{KFKI Research Institute for Particle and Nuclear Physics, Budapest, Hungary.}
\author{M.~Szuba}\affiliation{Faculty of Physics, Warsaw University of Technology, Warsaw, Poland.}
\author{P.~Szymanski}\affiliation{CERN, Geneva, Switzerland.}\affiliation{Institute for Nuclear Studies, Warsaw, Poland.}
\author{V.~Trubnikov}\affiliation{Institute for Nuclear Studies, Warsaw, Poland.}
\author{D.~Varga}\affiliation{KFKI Research Institute for Particle and Nuclear Physics, Budapest, Hungary.}\affiliation{CERN, Geneva, Switzerland.}
\author{M.~Vassiliou}\affiliation{Department of Physics, University of Athens, Athens, Greece.}
\author{G.I.~Veres}\affiliation{KFKI Research Institute for Particle and Nuclear Physics, Budapest, Hungary.}\affiliation{MIT, Cambridge, USA.}
\author{G.~Vesztergombi}\affiliation{KFKI Research Institute for Particle and Nuclear Physics, Budapest, Hungary.}\affiliation{KFKI Research Institute for Particle and Nuclear Physics, Budapest, Hungary.}
\author{D.~Vrani\'{c}}\affiliation{Gesellschaft f\"{u}r Schwerionenforschung (GSI), Darmstadt, Germany.}
\author{A.~Wetzler}\affiliation{Fachbereich Physik der Universit\"{a}t, Frankfurt, Germany.}
\author{Z.~W{\l}odarczyk}\affiliation{Institute of Physics \'Swi\c{e}tokrzyska Academy, Kielce, Poland.}
\author{A.~Wojtaszek}\affiliation{Institute of Physics \'Swi\c{e}tokrzyska Academy, Kielce, Poland.}
\author{I.K.~Yoo}\affiliation{Department of Physics, Pusan National University, Pusan, Republic of Korea.}
\author{J.~Zim\'{a}nyi}~\email[Deceased.]{}\affiliation{KFKI Research Institute for Particle and Nuclear Physics, Budapest, Hungary.}


\collaboration{The NA49 collaboration} \noaffiliation



\begin{abstract}

Transverse momentum spectra up to $4.5\GeVpc$ around midrapidity of 
$\pi^{\pm}$, $p$, $\bar{p}$, $K^{\pm}$ in Pb+Pb reactions were measured 
at $\sqrt{s_{{}_{NN}}}=17.3\GeV$ by the CERN-NA49 experiment. 
The nuclear modification factors $R_{AA}$ for $\pi^{\pm}$ 
and $R_{CP}$ for $\pi^{\pm},p,\bar{p},K^{\pm}$ were extracted and are compared to RHIC results at 
$\sqrt{s_{{}_{NN}}}=200\GeV$. The modification factor $R_{AA}$ shows a 
rapid increase with transverse momentum in the covered region. 
This indicates that the Cronin effect is the dominating effect in our 
energy range. The modification factor $R_{CP}$, 
in which the contribution of the Cronin effect is reduced,
shows a saturation 
well below unity in the $\pi^{\pm}$ channel. The extracted $R_{CP}$ values 
follow the $200\GeV$ RHIC results closely in the available transverse momentum 
range, except for $\pi^{\pm}$ above $2.5\GeV/c$ transverse momentum.
There the measured suppression is smaller than that observed at RHIC.

\end{abstract}


\maketitle


\section{Introduction}

One of the most interesting features discovered at RHIC
is the suppression of particle production at high transverse momenta in central 
nucleus-nucleus reactions relative to peripheral ones as well as to p+nucleus and to p+p collisions 
\cite{phenixauaunopid,phenixauaupid,phenixdaunopid,phenixdaupid,starauaupid}.
This is generally interpreted as a sign of parton energy loss in hot and dense strongly interacting 
matter.

The aim of the presented analysis is to investigate the energy dependence of these effects 
via a systematic study of Pb+Pb reactions at top ion-SPS energy, $158A\GeV$
($\sqrt{s_{{}_{NN}}}=17.3\GeV$), with the CERN-NA49 detector. 
A similar study has been published by the CERN-WA98 collaboration 
for the $\pi^{0}$ channel \cite{wa98}. The aim of this paper is to 
extend their results to all charged particle channels, i.e. $\pi^{\pm}$, $p$, $\bar{p}$ and $K^{\pm}$.

Invariant yields were extracted as a function of transverse momentum $p_{{}_T}$ in the 
range from $0.3$ to $4.5\GeVpc$ 
in the rapidity interval $-0.3\leq y\leq 0.7$ (midrapidity), at 
different collision centralities. 
The identification of particle types is crucial, because the particle 
composition of hadron spectra changes rapidly with transverse momentum and
differs significantly from that observed at RHIC energies 
(at SPS energies $p$ production is comparable to the $\pi^{\pm}$ yields already
at moderate values of $p_{{}_T}$).

The most important results are the nuclear modification factors $R_{AA}$, which were extracted 
from the identified charged hadron spectra in Pb+Pb and from 
the already published p+p pion spectra \cite{pp}. It is important 
to note that the use of reference spectra (e.g.\ p+p), measured exactly at the 
same collision energy is necessary, because of the rapid change of 
the shape of the particle spectra around $\sqrt{s_{{}_{NN}}}=17.3\GeV$. 
A baseline, which is constructed from the existing nearby energy measurements 
is not sufficient, as discussed in \cite{enterria}.

Besides the in-medium parton energy loss, the $R_{AA}$ quantity can also 
contain other medium effects, such as the \emph{Cronin effect} \cite{cronin}, 
a term used for the observation of an increased particle yield 
at high transverse momentum in p+nucleus compared to the elementary p+p reactions. 
This effect is believed to be a consequence of multiple scattering of projectile 
partons (or hadrons) in the nuclei before the particle production process. 
The Cronin effect could cause a large part of the nuclear modification 
effects measured by the $R_{AA}$ quantity. 
For a more sensitive search of the other nuclear effects 
the $R_{CP}$ (central to peripheral) modification factors were also extracted 
from the data. If the multiple scattering interpretation of the Cronin effect 
is correct, its contribution to the $R_{CP}$ ratio is expected to be reduced, as 
it should already be present in the peripheral baseline.

\section{Experimental Setup}

\begin{figure*}[!ht]
\includegraphics[width=16cm]{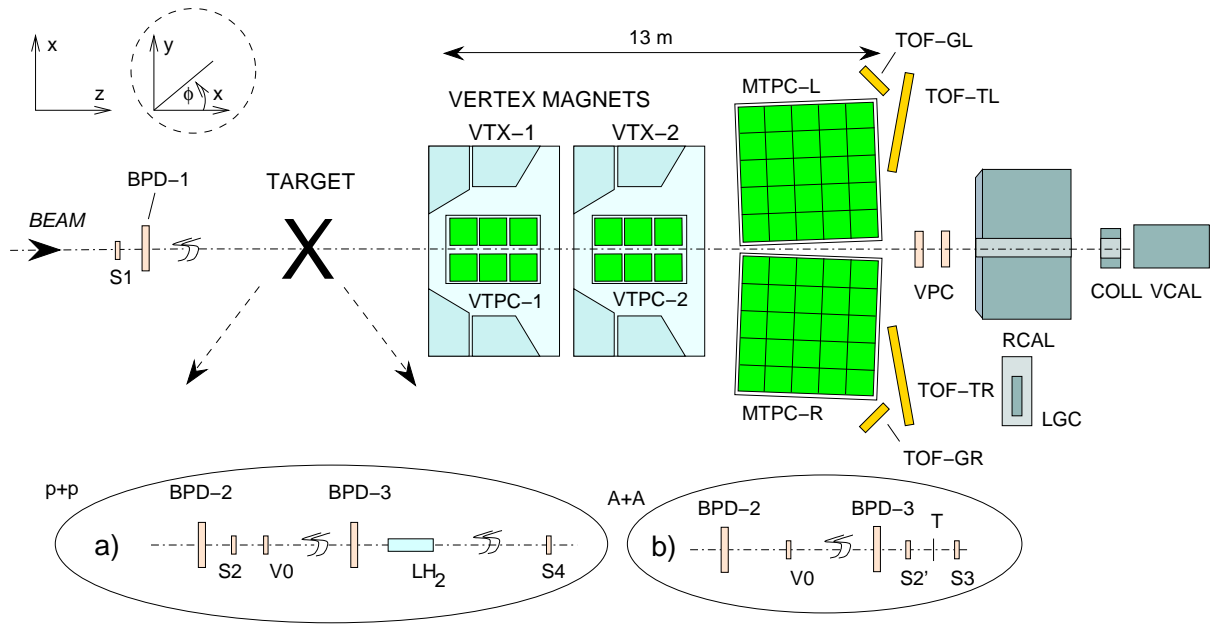}
\caption{(Color online) Setup of the CERN-NA49 experiment showing 
different beam definitions and target arrangements.}
\label{na49_setup}
\end{figure*}

The NA49 detector \cite{na49_nim} is a wide-acceptance hadron
spectrometer for the study of hadron production in collisions of
hadrons or heavy ions at the CERN-SPS. The main components 
are four large-volume Time Projection Chambers, TPCs, 
(Fig.\ \ref{na49_setup}) which are capable of detecting $80\%$ of
some $1500$ charged particles created in a central Pb+Pb collision
at $158A\GeV$ beam energy. Two chambers, the Vertex TPCs (VTPC-1 and
VTPC-2), are located in the magnetic field of two superconducting
dipole magnets ($1.5$ and $1.1\,\mathrm{T}$), while the two others
(MTPC-L and MTPC-R) are positioned downstream of the magnets
symmetrically to the beam line. The setup is supplemented by two
Time of Flight (TOF) detector arrays, which are not used in this 
analysis, and a set of calorimeters. The NA49 TPCs allow precise 
measurements of particle momenta $p$ with a resolution of 
$\sigma(p)/p^2 \cong(0.3-7)\cdot10^{-4}\,\mathrm{(GeV/c)^{-1}}$.

Pb foils (typically of $224\,\mathrm{mg/cm^2}$ thickness) are used as target for Pb+Pb
collisions, and a 
liquid hydrogen cylinder (length $20\,\mathrm{cm}$) for p+p interactions. 
The target is positioned about $80\,\mathrm{cm}$ upstream from VTPC-1.

Pb beam particles are identified by means of their charge as seen
by a Helium Gas-\v Cerenkov counter (S2') and proton beam particles 
by a $2\,\mathrm{mm}$ thick scintillator (S2). Both detectors are situated in 
front of the target. For p beams, interactions in the 
target are selected by requiring a valid incoming beam particle 
and no signal from a small scintillation counter (S4) placed on the beam line 
between the two vertex magnets. For p+p interactions at $158\GeV$ 
this counter selects a (trigger) cross section of $28.5\,\mathrm{mb}$ out of $31.6\,\mathrm{mb}$
of the total inelastic cross section. For Pb-ion beams an
interaction trigger is provided by a Helium
Gas-\v Cerenkov counter (S3) directly behind the target. The S3
counter is used to select minimum-bias collisions by requiring a
reduction of the \v Cerenkov signal. Since
the \v Cerenkov signal is proportional to $Z^2$, this requirement
ensures that the Pb projectile has interacted with a minimal
constraint on the type of interaction. This setup limits the
triggers on non-target interactions to rare beam-gas collisions,
the fraction of which proved to be small after cuts, even in the
case of peripheral Pb+Pb collisions. The resulting minimum-bias trigger 
cross section was about $80\%$ of the total inelastic cross section $\sigma_{\mathrm{Inel}}=7.15\,\mathrm{b}$.

For Pb+Pb reactions, the centrality of a collision is selected 
(on-line for central Pb+Pb, off-line for minimum-bias Pb+Pb interactions) by a 
trigger using information from a downstream calorimeter (VCAL), which 
measures the energy of the projectile spectator nucleons \cite{VCAL,spectpaper}.

\section{Analysis}

Table\ \ref{data} lists the statistics used in the $158A\GeV$ Pb+Pb collision analysis.

\begin{table}[!ht]
\begin{tabular}{|c|c|c|c|}
\hline
Reaction & $\sqrt{s_{{}_{NN}}}$ [GeV] & Centrality & Number of events\\
\hline
\hline
 Pb+Pb & $17.3$ & central & $830\,\mathrm{k}$ \\
\hline
 Pb+Pb & $17.3$ & mid-central & $1.4\,\mathrm{M}$ \\
\hline
 Pb+Pb & $17.3$ & peripheral & $200\,\mathrm{k}$ \\
\hline
\end{tabular}
\caption{The event statistics, used in the analysis.}
\label{data}
\end{table}

\subsection{Event Selection}

\begin{figure}[!ht]
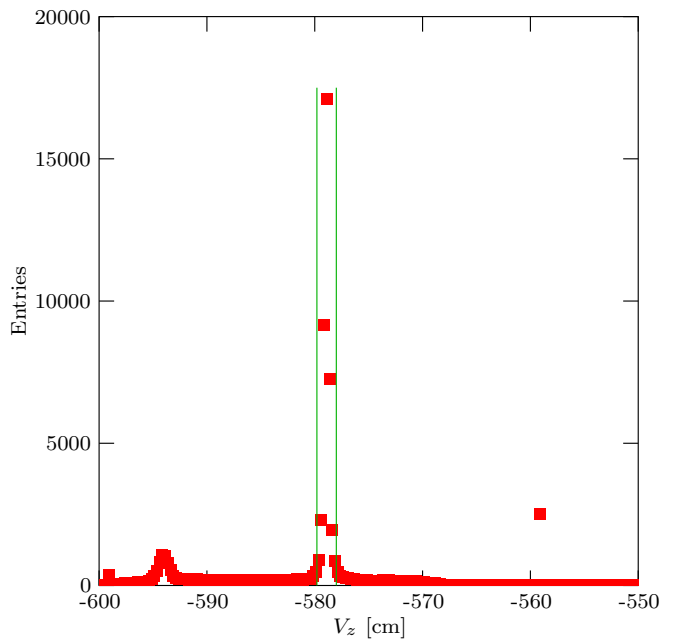

{\footnotesize\blopeps[width=8.5cm, height=8.5cm]{fig/Vz/Vz.beps}}
\caption{(Color online) The distribution of the longitudinal coordinate 
of the reconstructed collision point in minimum-bias Pb+Pb events, 
together with the cuts. The other peaks correspond to material in the beamline, 
while the small constant background under the target peak corresponds to 
beam+gas reactions.}
\label{Vz}
\end{figure}

As the target setup of the experiment is not contained in a vacuum pipe, 
non-target background collisions are also expected to be recorded in the 
unfiltered event sample. To reduce this contamination, a cut was applied 
to the reconstructed longitudinal position of the collision point: 
only those collisions were accepted, which occurred close 
to the nominal target position. An example plot is shown 
in Fig.\ \ref{Vz} for the case of Pb+Pb reactions, 
showing the target 
peak and the small flat non-target contribution, which was used to 
estimate the remaining background events. The contamination of the peripheral 
Pb+Pb spectra remains below $5\%$. For non-peripheral spectra this contamination 
is zero. The peripheral Pb+Pb spectra are corrected for this effect.

\begin{figure}[!ht]
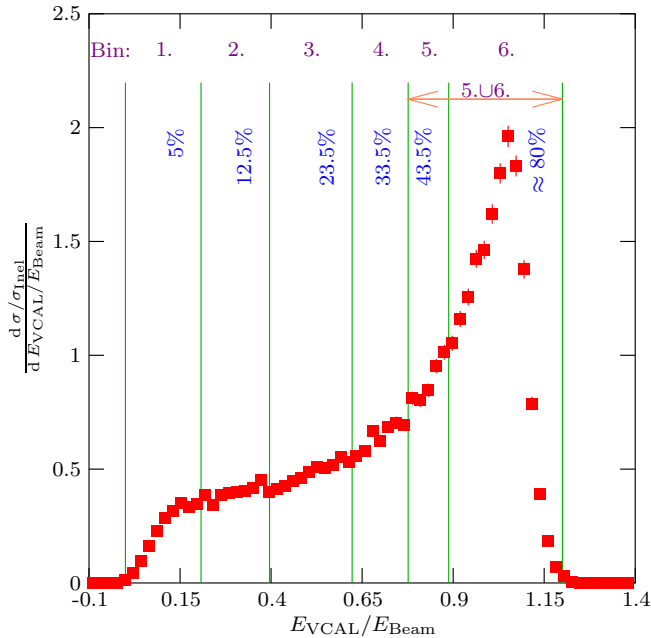

{\footnotesize\blopeps[width=8.5cm, height=8.5cm]{fig/VCAL/VCAL.beps}}
\caption{(Color online) The distribution of the energy deposited in the VCAL 
by the projectile spectators in Pb+Pb minimum-bias reactions. The centrality 
intervals selected by fraction of total inelastic cross section are shown 
by the vertical lines.}
\label{VCAL}
\end{figure}

In the case of Pb+Pb reactions, the recorded events were classified by 
collision centrality, which is correlated to the impact parameter -- 
the transverse distance of the centers of the colliding nuclei at impact. The 
centrality is determined via the energy measured by the VCAL:
the non-interacting (spectator) part of the projectile nucleus travels 
along the beamline, 
finally hitting the VCAL's surface, and leaving an energy signal 
in the apparatus, proportional to the volume of the projectile spectator 
matter. The higher this energy is, the more peripheral is the 
given reaction. The event centrality is defined by the fraction of total 
inelastic cross section, i.e.\ by the running integral
\[\frac{\sigma(E_{\mathrm{VCAL}})}{\sigma_{\mathrm{Inel}}}=\frac{1}{\sigma_{\mathrm{Inel}}}\int_{0}^{E_{\mathrm{VCAL}}}\frac{\mathrm{d}\sigma(E_{\mathrm{VCAL}}^{'})}{\mathrm{d}E_{\mathrm{VCAL}}^{'}}\;\mathrm{d}E_{\mathrm{VCAL}}^{'},\]
where $\sigma_{\mathrm{Inel}}$ is the total inelastic cross section, and 
$\frac{\mathrm{d}\sigma}{\mathrm{d}E_{\mathrm{VCAL}}}$ 
is the differential cross section in $E_{\mathrm{VCAL}}$, which is defined 
by the $E_{\mathrm{VCAL}}$ spectrum, normalizing its area to the trigger cross section. 
The VCAL energy spectrum and event centrality classification is shown in 
Fig.\ \ref{VCAL}. A relatively large 
centrality interval was used for the selection of peripheral events in 
order to gain enough statistics in the particle spectrum analysis.

Due to the aging of the photomultiplier tubes of the calorimeter, the VCAL response was slowly varying 
in time in terms of scale offset and amplification. 
This effect is corrected by a time-dependence calibration procedure, 
which is based on keeping the correlation of the VCAL energy 
to the total measured charged multiplicity time-independent (see \cite{VCALrel}).
The mean values of various collision parameters 
(number of wounded nucleons, number of binary collisions) 
for the selected centrality intervals were calculated from the VENUS-4.12 Monte Carlo model \cite{venus}
which was filtered by the detector simulation. 
For the non-peripheral region, this was straightforward by applying 
statistical averaging in the centrality intervals. For the peripheral 
region, a semi-empiric approach was used, which also takes the trigger bias into account: 
the VCAL energy distribution for the averaging was taken from the measurement, 
while the mapping of VCAL energy into collision parameters was taken from 
VENUS + detector simulation. The procedure is described in \cite{VCALabs}. A 
list of the average values is shown in Table\ \ref{avepbpb} together with their 
systematic errors, which were estimated by assuming a $5\%$ 
uncertainty of the total inelastic cross section, and a $10\%$ uncertainty 
of the skin thickness of the nuclear density profile. The stated values were 
also confirmed by the GLISSANDO Monte Carlo model \cite{glissando}. 

\begin{table}[!ht]
\begin{tabular}{|c|c|c|c|}
\hline
Bin & Centrality & $\left<N_{W}\right>$ & $\left<N_{{}_{BC}}\right>$\\
\hline
\hline
 1. & $0-5\%$ & $357\pm2$ & $742\pm18$\\
\hline
 2. & $5-12.5\%$ & $288\pm3$ & $565\pm18$\\
\hline
 3. & $12.5-23.5\%$ & $211\pm4$ & $379\pm16$\\
\hline
 4. & $23.5-33.5\%$ & $146\pm5$ & $234\pm14$\\
\hline
 5. & $33.5-43.5\%$ & $87 \pm7$$\;{}^{\star}$ & $120\pm13$$\;{}^{\star}$\\
\hline
 6. & $43.5-80\%$ & $40 \pm4$$\;{}^{\star}$ & $46 \pm5$$\;{}^{\star}$\\
\hline
 5.$\cup$6. & $33.5-80\%$ & $56 \pm7$$\;{}^{\star}$ & $70\pm13$$\;{}^{\star}$\\
\hline
\end{tabular}
\caption{The mean values of the number of wounded nucleons and binary collisions 
in various centrality intervals in Pb+Pb collisions, together with their systematic errors. 
${}^{\star}$: Averages for the full minimum-bias dataset using the semi-empiric method, as discussed 
in the text.}
\label{avepbpb}
\end{table}

\subsection{Track Selection}

The transverse momentum spectra of particles in heavy-ion collisions are 
known to decrease rapidly (approximately exponentially) toward higher 
transverse momenta. Therefore, the background of fake tracks 
becomes more and 
more important with increasing $p_{{}_T}$, especially for fixed-target experiments, 
in which the track density is strongly focused in the forward direction, 
increasing the probability for fake track formation. 
We apply cuts to minimize the contribution of such artificial tracks. 
These background tracks consist of discontinuous track candidates, and track 
candidates whose trajectories pass close to the border of 
the detector volume. The former kind of tracks mainly arise from erroneous pairing 
of straight track pieces in MTPC, outside of the magnetic field, to some residual 
points in the VTPCs. The latter kind of tracks are subject to distortions, as they are 
likely to have missing or displaced 
clusters. The corresponding trajectories will have wrong fitted curvature, 
therefore they should be discarded from the analysis.

\begin{figure}[!ht]
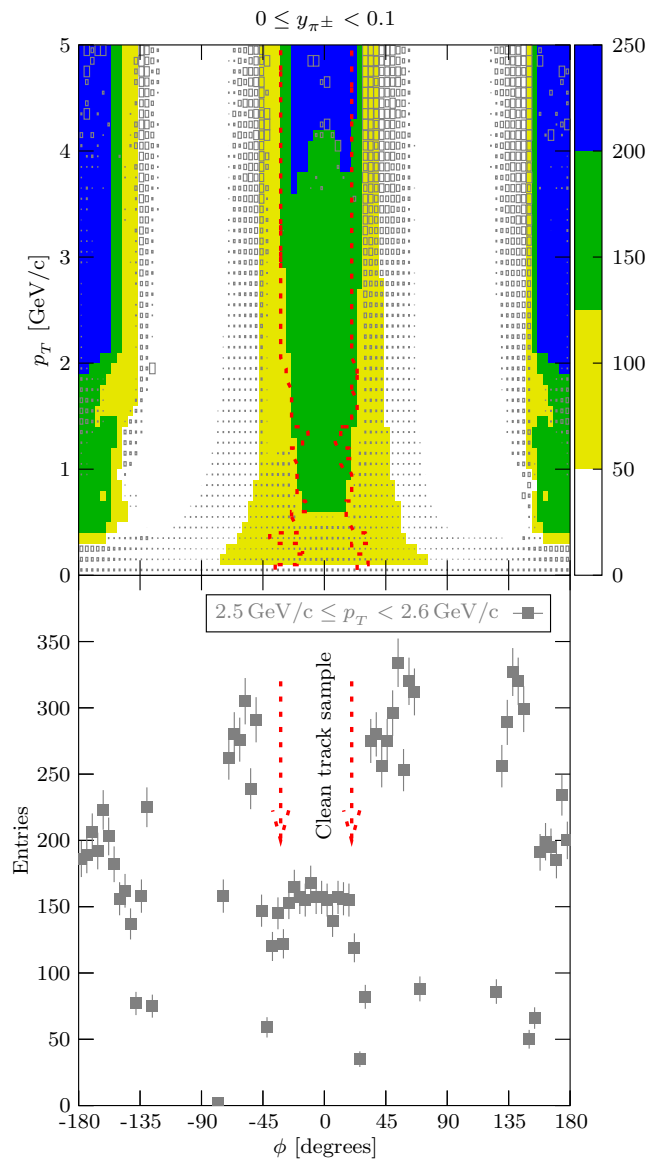

{\footnotesize\blopeps[width=8.5cm, height=15.5cm]{fig/quality/Qual.beps}}
\caption{(Color online) Upper panel: example plot for the momentum space selection 
in the rapidity slice $0\leq y<0.1$, with $\pi^{\pm}$ mass hypothesis. 
$\phi$ means charge-reflected azimuth.
The size of the boxes indicate the fraction of those track candidates in the 
given momentum space bin, which have a ratio of number of measured 
to number of potential points 
below $60\%$ (i.e.\ the fraction of those track candidates, 
which most likely do not correspond to real particle trajectories).
The color map indicates the number of potential points as a function of momentum space. 
As can be seen, the bad track candidates mainly populate the borders of the acceptance. 
The dotted line shows the momentum space cut, guided by potential point isosurfaces.
Lower panel: example plot for the $\phi$ distribution at $0\leq y<0.1$ and $2.5\GeVpc\leq p_{{}_T}<2.6\GeVpc$.}
\label{hipt}
\end{figure}

The discontinuous track
candidates can be easily identified: they are track candidates that
(according to their reconstructed charge and momentum) would have left hits
in a given TPC (VTPC1, VTPC2, or MTPC), but did not.
These tracks were rejected as 
one source of fake tracks. 
After this cleaning procedure a 3 dimensional phase space study 
was performed. Its coordinates are rapidity $y$, transverse momentum $p_{{}_T}$ 
and charge-reflected azimuth $\phi$. Charge-reflected azimuth is defined by the azimuth 
angle of the momentum vector in the transverse plane, whose $x$-component is 
multiplied by the charge sign of the track candidate -- the $x$-direction being 
perpendicular to the magnetic field and to the beamline. As our detector has 
an $x$-reflection symmetry, the use of charge-reflected azimuth enables us to 
distinguish tracks not crossing the plane defined by the beamline and the 
magnetic field direction (right-side tracks) from tracks crossing this plane 
(wrong-side tracks). An example of the track quality distribution as a function of $\phi$ and $p_{{}_T}$
is shown in the upper panel of Fig.\ \ref{hipt} for a typical rapidity slice. 
A domain of high quality right-side tracks reveals 
itself: in this region, the fraction of those tracks with a ratio of the 
number of measured to number of potential points below $60\%$ is very low.
Selection of this 
region by a 3 dimensional momentum space cut provides a clean track sample. 
The 3 dimensional momentum 
space cut is guided by the isosurfaces of the number of potential points, 
and by the requirement of avoiding efficiency holes, which can be determined 
from the dropping of the $\phi$ distribution at fixed $(y,p_{{}_T})$ values 
(the $\phi$ distribution should be uniform due to the axial symmetry of 
particle production). 
The momentum space cut curve is shown by the dotted line in the upper panel 
of Fig.\ \ref{hipt}, at given $y$. The $\phi$ distribution is shown in 
the lower panel of Fig.\ \ref{hipt} at given $(y,p_{{}_T})$, which reveals 
the efficiency holes to be avoided, and the high track yield excess at 
the acceptance borders. The dotted arrows indicate the momentum space cut in 
$\phi$, for this particular $(y,p_{{}_T})$ slice. The outlined procedure 
is discussed in detail in \cite{hipttech}. The resulting phase space, which 
contains the clean track sample, 
covers the rapidity region $-0.3\leq y<0.7$, the transverse momentum 
region $0\GeVpc\leq p_{{}_T}<5\GeVpc$, and a $(y,p_{{}_T})$ dependent 
$\phi$ interval. In the following analysis, all distributions will be shown 
as a function of transverse momentum.

\begin{figure}[!ht]
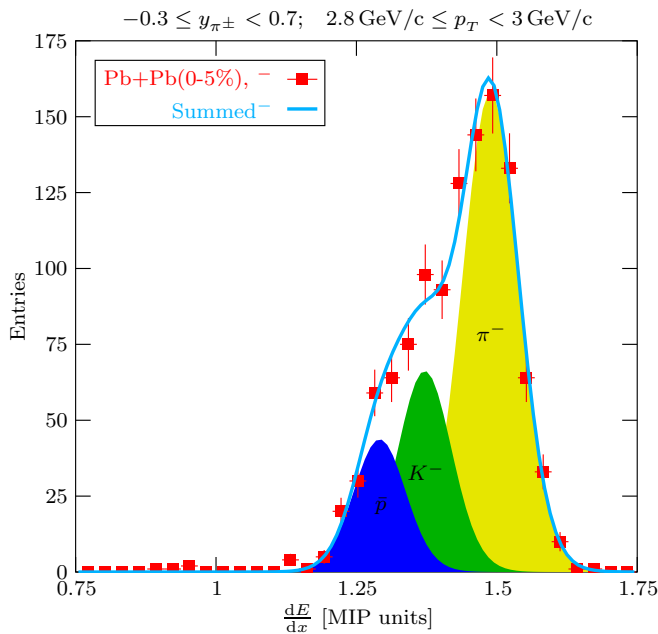

{\footnotesize\blopeps[width=8.5cm, height=8.5cm]{fig/pid/PID.beps}}
\caption{(Color online) Example plot for the identification of negative particles
by specific energy loss fits in central Pb+Pb collisions (0-5\%). 
The parameterization of the response function of 
each particle species with fixed momentum is known, but the yields 
(the amplitudes in the summed $\frac{\mathrm{d}E}{\mathrm{d}x}$ spectrum model) 
are fit parameters.}
\label{dEdx}
\end{figure}

The particle identification was performed via the specific energy loss 
($\frac{\mathrm{d}E}{\mathrm{d}x}$) of the particles. In each transverse 
momentum bin the $\frac{\mathrm{d}E}{\mathrm{d}x}$ spectrum was recorded, 
and by using the known shape of the response function (see \cite{gabordEdx,dezsodEdx}) 
of the given particle species ($\pi^{\pm}$, $p,\bar{p}$, $K^{\pm}$, $e^{\pm}$), 
a fit was performed to the $\frac{\mathrm{d}E}{\mathrm{d}x}$ histogram, 
with the amplitudes of the particle response functions 
as free parameters. Also other parameters characteristic of the response function 
shape, such as the most probable value, were kept as free parameters. 
Thus, their statistical errors are also reflected in the fitted errors of the amplitudes.
As our data was limited by statistics at higher $p_{{}_T}$, 
Poisson maximum-likelihood fitting was employed. This approach allows reliable 
fits even in regions with low statistics (see \cite{poisson}).
An example can be seen in Fig.\ \ref{dEdx} which demonstrates 
the $\frac{\mathrm{d}E}{\mathrm{d}x}$ fit procedure for inclusive 
particle identification in a typical phase space bin. The full procedure is 
described in \cite{hipttech}. Systematic errors, caused by possible 
systematic shifts of fit parameters were estimated by calculating the statistical 
regression matrix from the fitted covariance matrix, and by propagating 
the systematic errors of the fit parameters (approximately estimated to be about 
$0.5\%$) to the fitted yields via this matrix. The resulting systematic errors 
of yields are estimated to be $4\%$ for $K^{+}$, $2\%$ for $\bar{p}$, and $1\%$ for 
$\pi^{+}$, $\pi^{-}$, $p$, $K^{-}$.

\subsection{Corrections}

Monte Carlo methods including a GEANT simulation of our 
detector were used to correct the yields of $\pi^{+},\pi^{-}$ and $K^{+},K^{-}$ 
as well as $p$ and $\bar{p}$ for tracking inefficiencies (below $10\%$), decay losses 
(from $20\%$ to $0\%$), feed-down from weakly decaying particles ($5-30\%$), 
and finally for geometric acceptance. The peripheral particle spectra were also 
corrected for non-target contamination ($5\%$) by extracting the non-target 
contribution from pure beam+gas events, and subtracting it 
(for detailed procedure, see \cite{hipttech}). 
The yield of fake tracks, after the 
introduced track cuts, turned out to be negligible. The momentum space 
resolution was estimated to be better than $1\%$ in the whole region. The momentum 
scale uncertainty is of the order of $0.1\%$.

The largest and most sensitive 
correction is the feed-down correction. The yield of feed-down particles 
was estimated by using a simulation to determine the conditional probabilities 
of reconstructing a secondary particle originating from a weakly decaying primary particle 
as a primary particle. 
To calculate the feed-down yields as a function of momentum, one has to fold these tables of 
conditional probabilities with the momentum differentiated yields of the weakly decaying particles. 
The relevant decay channels are listed in Table\ \ref{fdchan}. 
The yields of the weakly decaying particles $K^{0}_{s}$, $\Lambda,\bar{\Lambda}$ 
were taken from parameterizations of yields, measured previously in our experiment 
\cite{k0s,lambda,compil} ($K^{0}_{s}$ yield was constructed from average charged Kaon data).
The used $\Lambda,\bar{\Lambda}$ yields include the 
$\Sigma^{0}\rightarrow \Lambda \gamma$ and the $\bar{\Sigma}^{0}\rightarrow \bar{\Lambda} \gamma$ contributions, 
but are feed-down corrected for $\Xi$ ($\bar{\Xi}$) decays. 
Since the $\Xi$ ($\bar{\Xi}$) yields are generally below $20\%$ of the $\Lambda$ ($\bar{\Lambda}$) 
yields, their contribution is neglected in the correction. However, it constitutes a source of 
systematic errors, that is discussed below.
The $K^{0}_{s}$ yield does not suffer from feed-down contamination.

\begin{table}[!ht]
\begin{tabular}{lcll}
 $K^{0}_{s}$        & $\rightarrow$ & $\pi^{+}\;\pi^{-}$ & $\quad\leq5\%\;\,$ to $\pi^{\pm}$ \\
 $\Lambda$          & $\rightarrow$ & $p\;\pi^{-}$ & $\quad\leq30\%$ to $p$ \\
 $\bar{\Lambda}$    & $\rightarrow$ & $\bar{p}\;\pi^{+}$ & $\quad\leq30\%$ to $\bar{p}$ \\
 $\Sigma^{+}$       & $\rightarrow$ & $p\;\pi^{0}$ & $\quad\leq6\%\;\,$ to $p$ \\
 $\Sigma^{+}$       & $\rightarrow$ & $n\;\pi^{+}$ & $\quad$negligible \\
 $\Sigma^{-}$       & $\rightarrow$ & $n\;\pi^{-}$ & $\quad$negligible \\
 $\bar{\Sigma}^{-}$ & $\rightarrow$ & $\bar{p}\;\pi^{0}$ & $\quad\leq6\%\;\,$ to $\bar{p}$ \\
 $\bar{\Sigma}^{-}$ & $\rightarrow$ & $\bar{n}\;\pi^{-}$ & $\quad$negligible \\
 $\bar{\Sigma}^{+}$ & $\rightarrow$ & $\bar{n}\;\pi^{+}$ & $\quad$negligible \\
\end{tabular}
\caption{The list of relevant feed-down channels.}
\label{fdchan}
\end{table}

\begin{figure}[!ht]
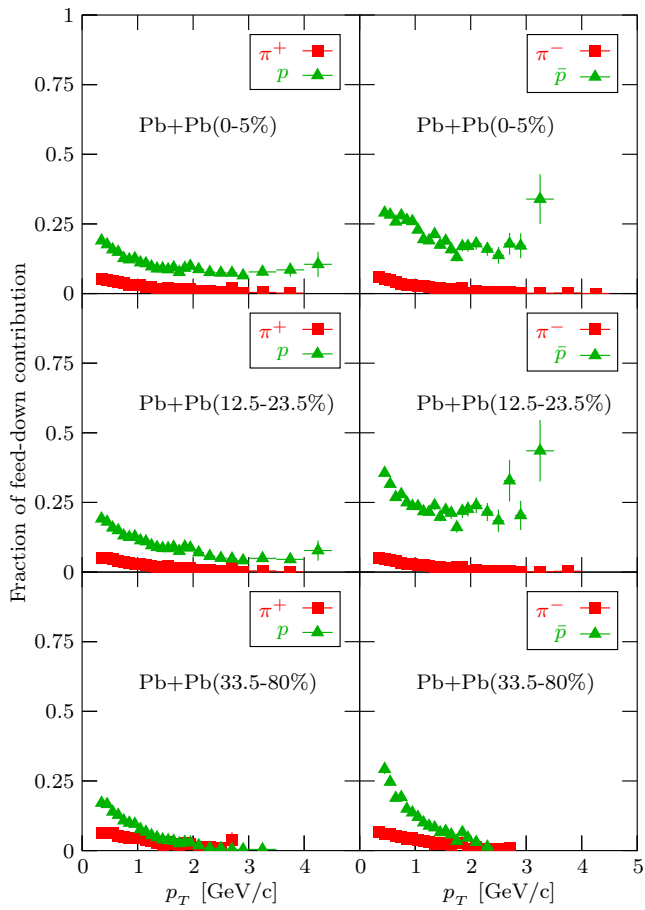

{\footnotesize\blopeps[width=8.5cm, height=12.25cm]{fig/fd/fd.beps}}
\caption{(Color online) Ratio of the estimated feed-down contribution to the raw measured
particle yields. The top, center and bottom row of plots correspond, respectively, to Pb+Pb
collisions with centrality ranges of
0-5\%, 12.5-23.5\% and 33.5-80\% of the inelastic cross section (see Table \ref{avepbpb}).}
\label{feeddown}
\end{figure}

For $\pi^{+}$, only the $K^{0}_{s}\rightarrow \pi^{+} \pi^{-}$ channel gives a
sizable contribution, while for $\pi^{-}$ also the $\Lambda\rightarrow p \pi^{-}$ channel 
has to be taken into account. All other contributions to the $\pi^{\pm}$ channels are negligible. 
For $p$, the $\Lambda\rightarrow p \pi^{-}$ gives 
the dominant contribution, and for $\bar{p}$, the $\bar{\Lambda}\rightarrow \bar{p} \pi^{+}$ 
is dominant. The $p$ and $\bar{p}$ spectra are also contaminated by the decays of 
$\Sigma^{+}$ and $\bar{\Sigma}^{-}$. These are taken into account 
by scaling up the $\Lambda,\bar{\Lambda}$ yields by $20\%$. This treatment was 
suggested by the VENUS-4.12 model, which predicts that 
the relative intensity of the non-$\Lambda$ contribution 
to the $p$ feed-down, or the relative intensity of the 
non-$\bar{\Lambda}$ contribution to the $\bar{p}$ 
feed-down is approximately constant at $20\%$. 
Assuming a $50\%$ systematic uncertainty of this $20\%$ scaling, the 
contribution to the systematic uncertainty of the $p,\bar{p}$ spectra would be $3\%$.
There is an additional uncertainty of the $\bar{p}$ feed-down correction, 
which is due to the large errors of the measured $\bar{\Lambda}$ inverse slope parameters. 
The systematic error, caused by the poor knowledge of the $\bar{\Lambda}$ 
slopes was estimated by repeating the feed-down calculation using 
the $\Lambda$ temperatures, which should provide a reasonable upper bound for the 
$\bar{\Lambda}$ temperatures. The systematic error, caused by this variation 
was estimated to be $5\%$.
The calculated feed-down contribution to the $\pi^{\pm}$ and $p,\bar{p}$ yields in 
Pb+Pb reactions is depicted in Fig.\ \ref{feeddown}.

\begin{figure}[!ht]
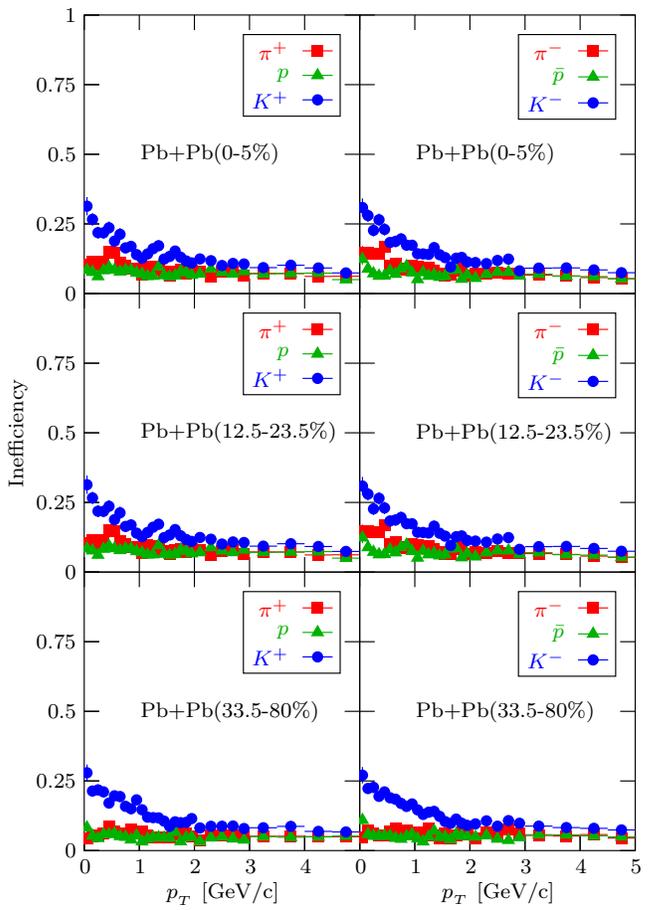

{\footnotesize\blopeps[width=8.5cm, height=12.25cm]{fig/ineff/ineff.beps}}
\caption{(Color online) Detection inefficiencies for 
$\pi^{\pm}$ (squares), $K^{\pm}$ (circles),
$p$ and $\bar{p}$ (triangles). The top, center and bottom row of plots
correspond to Pb+Pb collisions with centrality ranges of
0-5\%, 12.5-23.5\% and 33.5-80\% of the inelastic cross section (see
Table \ref{avepbpb}). The decay loss for $\pi^{\pm}$ and $K^{\pm}$ 
is included in the inefficiencies.}
\label{ineff}
\end{figure}

The tracking inefficiency was estimated by simulation and reconstruction of 
embedded particles in real events, i.e.\ by determining the conditional 
probability of losing a track of given particle type and momentum, taking 
the proper track density environment into account. The inefficiency estimates are 
plotted in Fig.\ \ref{ineff}. These also include the inefficiency 
caused by the decay loss for the weakly decaying particles.

The aim of our analysis is to measure the particle yields as a function 
of $p_{{}_T}$ at $y=0$, integrated over $-180^{\circ}\leq\phi<180^{\circ}$. 
Therefore, the particle spectra have to be corrected for the geometric 
acceptance. This correction can be performed without simulation, 
as the accepted momentum space region is defined explicitly by our 
momentum space cut: only the $(y,\phi)$ shape of the particle distributions have 
to be known in each $p_{{}_T}$ bin. 
The $\phi$ distribution is uniform in each $(y,p_{{}_T})$ slice due to the 
axial symmetry of particle production. For the $y$-dependence we use the fact, 
that the $y$ spectra around midrapidity are approximately 
flat in Pb+Pb reactions (see \cite{pbpbold1}). Therefore, we 
assume a flat $y$ distribution in $-0.3\leq y<0.7$ for the acceptance 
correction. The systematic error, caused by this approximation has been 
estimated to be below $2\%$.

When all the discussed corrections are taken into account, the Pb+Pb particle 
spectra bear the systematic errors listed in Table \ref{syst}.

\begin{table}[!ht]
\begin{tabular}{l|l|l|l|l|l}
 particle    & $\frac{\mathrm{d}E}{\mathrm{d}x}$ & acceptance & feed-down& feed-down & quadratic \\
 type        &       & correction & yields   & shapes    & sum \\
\hline
 $\pi^{+}$   & $1\%$ & $2\%$ &       &       & $2.2\%$ \\
 $\pi^{-}$   & $1\%$ & $2\%$ &       &       & $2.2\%$ \\
 $p$         & $1\%$ & $2\%$ & $3\%$ &       & $3.7\%$ \\
 $\bar{p}$   & $2\%$ & $2\%$ & $3\%$ & $5\%$ & $6.5\%$ \\
 $K^{+}$     & $4\%$ & $2\%$ &       &       & $4.5\%$ \\
 $K^{-}$     & $1\%$ & $2\%$ &       &       & $2.2\%$ \\
\end{tabular}
\caption{The list of estimated systematic errors and their sources on the 
$\frac{\mathrm{d}n}{\mathrm{d}p_{{}_T}}$ values.}
\label{syst}
\end{table}

Points which correspond to a fitted amplitude of less than $25$ particles or 
with a statistical error above $30\%$ are not shown. 
Points below $p_{{}_T} = 0.3\GeVpc$ are not reported at all in order to avoid 
the $\frac{\mathrm{d}E}{\mathrm{d}x}$ cross-over region.

\section{Results}

\subsection{Charged Hadron Spectra}

\begin{figure}[!ht]
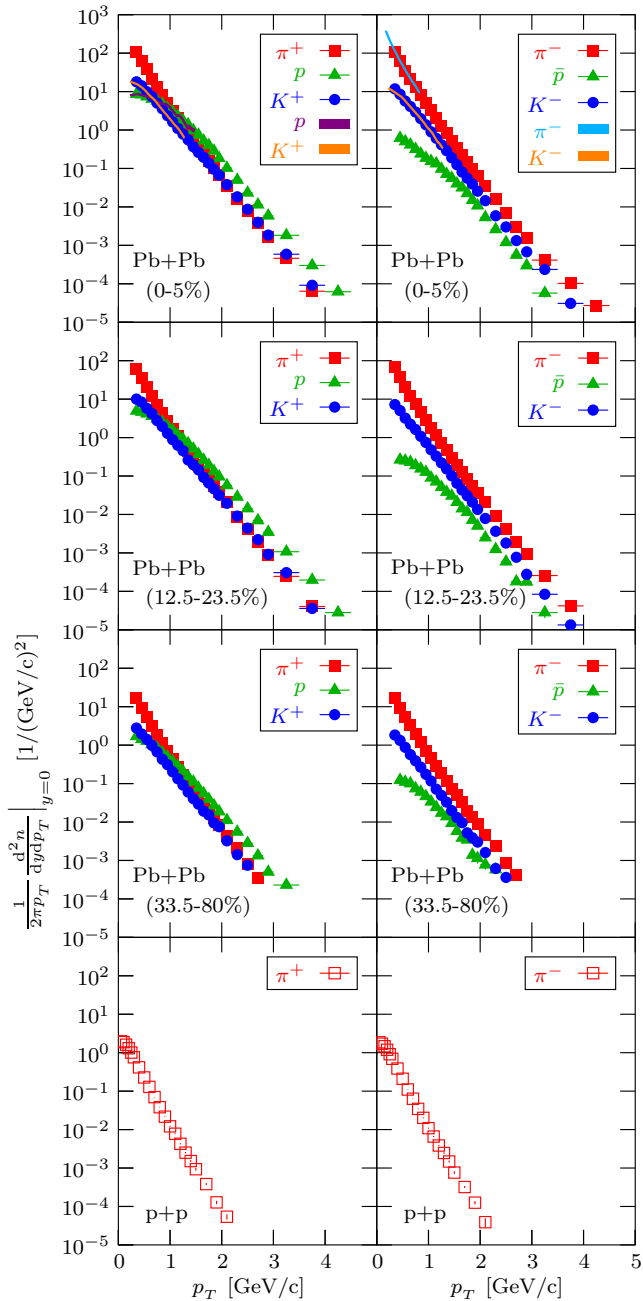

{\footnotesize\blopeps[width=8.5cm, height=17.5cm]{fig/spec/spec3.beps}}
\caption{(Color online) Invariant yields per inelastic collision of 
$\pi^{\pm}$ (squares), $K^{\pm}$ (circles) and 
$p,\bar{p}$ (triangles) versus transverse momentum $p_{{}_T}$ at midrapidity
and $\sqrt{s_{{}_{NN}}}=17.3\GeV$ energy. Results are shown for Pb+Pb collisions with
centrality ranges of 0-5\% (top row), 12.5-23.5\% (second row) and 33.5-80\% (third
row) of the inelastic cross section (see Table \ref{avepbpb}). Previously published
results for central collisions in the lower $p_{{}_T}$ range \cite{pbpbold1,pbpbold2} are indicated
by the bands in the top row. The bottom row shows
published NA49 results \cite{pp} for $\pi^{+}$ (left) and $\pi^{-}$ (right)
production in inelastic p+p collisions.}
\label{spectra}
\end{figure}

The resulting hadron spectra are shown in Fig.\ \ref{spectra}. 
Preliminary versions, without the discussed corrections have been shown 
in \cite{qm05,sqm05,qm06}. The new spectra show very good 
agreement with earlier low $p_{{}_T}$ Pb+Pb results \cite{pbpbold1,pbpbold2}. 
Also shown in Fig.\ \ref{spectra} (lowest panel) are our $\pi^{+},\pi^{-}$ spectra 
measured in p+p interactions \cite{pp}.

\subsection{Nuclear Modification Factors}

The nuclear modification effects are measured by the nuclear modification 
factors, which are particle yield ratios of two reactions, with  
appropriate scaling factors. The nuclear modification factors of a reaction 
$A_{1}+A_{2}$ relative to $A_{3}+A_{4}$ are most generally defined as:
\[R^{{}^{BC}}_{A_{1}+A_{2}/A_{3}+A_{4}}=\frac{\left<N_{{}_{BC}}(A_{3}+A_{4})\right>}{\left<N_{{}_{BC}}(A_{1}+A_{2})\right>}\cdot\frac{\mathrm{yield}(A_{1}+A_{2})}{\mathrm{yield}(A_{3}+A_{4})},\]
and 
\[R^{W}_{A_{1}+A_{2}/A_{3}+A_{4}}=\frac{\left<N_{W}(A_{3}+A_{4})\right>}{\left<N_{W}(A_{1}+A_{2})\right>}\cdot\frac{\mathrm{yield}(A_{1}+A_{2})}{\mathrm{yield}(A_{3}+A_{4})}.\]
The ratio $R_{A+A/p+p}$ is in the following abbreviated by the commonly used $R_{AA}$, while 
for $R_{p+A/p+p}$ the notation $R_{pA}$ will be used.
The scaling factors $\left<N_{{}_{BC}}\right>$ and $\left<N_{W}\right>$ are the average numbers of binary 
collisions and of wounded nucleons, respectively, which are calculated 
according to the Glauber model by Monte Carlo 
methods which simulate the geometrical configurations. 
The two different kinds of scaling assumptions 
are motivated by extreme scenarios of particle production. A pQCD-like 
particle production scheme would imply a binary collision scaling (particles 
are produced in binary parton-parton collisions). However, a scenario of a superposition
of soft collisions (e.g.\ particle production via projectile excitation and decay) would rather 
suggest the scaling of particle production with the number of wounded 
nucleons, as proposed in e.g.\ \cite{bialas} (each collision increases the energy content of the 
excited nucleon). As the particle production scheme has not been completely determined 
at SPS energies, especially below $2\GeVpc$ transverse momentum, both extreme 
scaling assumptions are examined.

\begin{figure}[!ht]
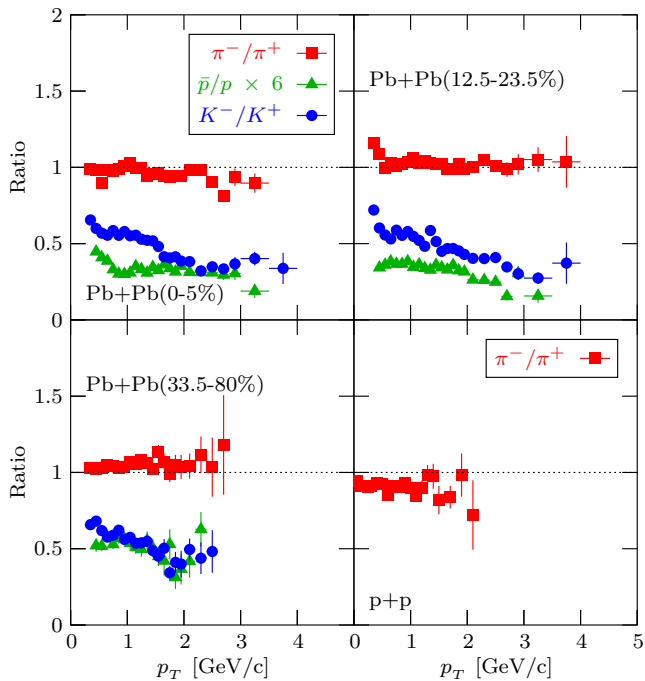

{\footnotesize\blopeps[width=8.5cm, height=9.25cm]{fig/r/asymm3.beps}}
\caption{(Color online) Ratios between antiparticle and particle yields versus transverse
momentum $p_{{}_T}$ at midrapidity and $\sqrt{s_{{}_{NN}}}=17.3\GeV$ energy. Results are
shown for $\pi^{-}/\pi^{+}$ (squares), $K^{-}/K^{+}$ (circles) and $\bar{p}/p$ (scaled up
by a factor $6$, triangles). Plots correspond to Pb+Pb collisions with
centrality ranges of 0-5\% (top left), 12.5-23.5\% (top right) and
33.5-80\% (bottom left) of the inelastic cross section (see
Table \ref{avepbpb}). Ratios from published NA49 results on $\pi^{-}$ and $\pi^{+}$
yields in p+p collisions \cite{pp} are plotted in the bottom right panel.}
\label{asymm}
\end{figure}

To verify the presence or absence of nuclear effects, we calculate the modification 
factors $R_{AA}$
and $R_{CP}$ (A+A central/peripheral) 
for hadrons using our particle spectra at $\sqrt{s_{{}_{NN}}}=17.3\GeV$. We compare 
our results to those obtained at $\sqrt{s_{{}_{NN}}}=200\GeV$, 
measured by the PHENIX experiment at RHIC \cite{phenixauaupid,phenixdaupid}. As the production of 
$\pi^{+}$ and $\pi^{-}$ is already very similar at our energy, we 
only show the average of $\pi^{\pm}$ for charged pions. 
The heavier particle yields ($p,\bar{p}$, $K^{+},K^{-}$) proved not yet to 
be symmetric with respect to charge conjugation at our energy and we show them separately. 
The antiparticle/particle asymmetry of our spectra can be seen in 
Fig.\ \ref{asymm}. As p+p is a maximally isospin-asymmetric 
system, the $\pi^{-}/\pi^{+}$ ratio is slightly below one. Pb+Pb 
is less asymmetric in isospin, thus its $\pi^{-}/\pi^{+}$ production ratio is closer to 
unity. The isospin argument does not hold for the case of $K^{-}/K^{+}$ and
$\bar{p}/p$ ratios, which are strongly influenced by the net-baryon density.

\begin{figure}[!ht]
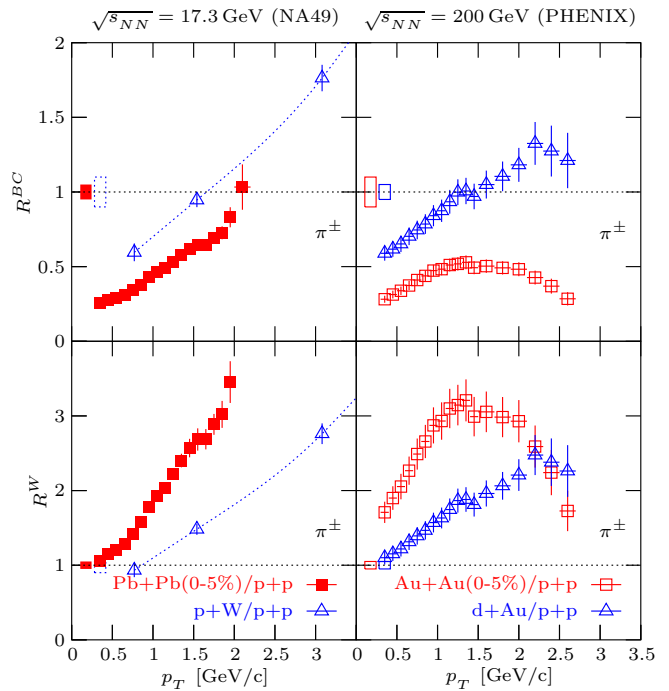

{\scriptsize\blopeps[width=8.5cm, height=9.25cm]{fig/r/Rbw6.beps}}
\caption{(Color online) Ratios of charged pion yields at midrapidity in central nucleus+nucleus and
inelastic nucleon+nucleon collisions scaled by the number of binary
collisions ($R^{BC}$, top row) and number of wounded nucleons ($R^{W}$, bottom row).
The left column
shows results from NA49 at the SPS ($\sqrt{s_{{}_{NN}}}=17.3\GeV$) for the
ratio of central (0-5\%) Pb+Pb collisions to inelastic p+p reactions
(filled squares) compared to the ratio of p+W to p+p collisions at
$\sqrt{s_{{}_{NN}}}=19.4\GeV$ \cite{cronin} (the dotted line is shown to guide the eye).
The right column shows measurements from RHIC ($\sqrt{s_{{}_{NN}}}=200\GeV$) \cite{phenixauaupid,phenixdaupid}
for the ratio of central (0-5\%) Au+Au collisions (open squares)
and d+Au reactions (open triangles). (The error bars attached to the
constant $1$ line indicate normalization uncertainty.)}
\label{r}
\end{figure}

The nuclear modification factor $R_{AA}$ 
is shown in Fig.\ \ref{r} for the $\pi^{\pm}$ channel at top SPS and RHIC collision energies. 
The RHIC data are compared to $R_{dA}$ at the same center-of-mass energy, while for the SPS 
data $R_{pA}$, measured at $\sqrt{s_{{}_{NN}}}=19.4\GeV$ \cite{cronin}, is shown in addition.
Since the $R_{pA}$ factors exhibit only a weak energy dependence for $0\leq p_{{}_T}<4.5\GeVpc$ 
in the energy range $19.4\leq\sqrt{s_{{}_{NN}}}\leq 27.4\GeV$ \cite{cronin}, 
the comparison to the $\sqrt{s_{{}_{NN}}}=17.3\GeV$ data is to a certain extent justified.
However, the shrinking of the available momentum space region with decreasing 
collision energy makes this extrapolation in energy somewhat uncertain, 
as the modification factors are expected to behave singularly close to the 
momentum space kinematic limit.

The main observations with the binary collision scaling 
assumption are the following. Both the $R_{AA}$ and the $R_{pA}$ 
data show a similar low $p_{{}_T}$ increase at $\sqrt{s_{{}_{NN}}}=17.3\GeV$ 
and $\sqrt{s_{{}_{NN}}}=200\GeV$, up to $p_{{}_T}\leq1\GeVpc$. 
The $R_{pA}$ ratios rise above $1$ at both collision energies 
with similar slopes, showing an excess of particle yield, which is often called 
Cronin effect. The $17.3\GeV$ $R_{AA}$ data keeps rising approximately linearly up to 
$2\GeVpc$, where the statistics of the p+p reference spectrum runs out. 
In the case of $200\GeV$ $R_{AA}$ data, 
the ratio starts to decrease in this region, showing a large 
suppression both relative to the constant $1$ line and to the $R_{pA}$ ratio 
(the ``Cronin baseline''). The evolution of the $17.3\GeV$ data at 
$p_{{}_T}>2\GeVpc$ is not clear, because the available experiments on p+p have no
statistics beyond $p_{{}_T}=2\GeVpc$. In the covered $p_{{}_T}$ region, 
the $R_{AA}$ ratio stays below the $R_{pA}$ Cronin baseline also 
at $17.3\GeV$.

When assuming wounded nucleon scaling, there are some differences compared to 
the above picture. 
Namely, the modification factors start from one at both energies, and show a rapid 
increase with similar slopes at the beginning. The $R_{pA}$ ratios stay 
below the $R_{AA}$ result as opposed to the binary collision case. 
The $200\GeV$ $R_{AA}$ data show a tendency 
of returning to the constant $1$ line at $p_{{}_T}\approx3\GeVpc$.

\begin{figure}[!ht]
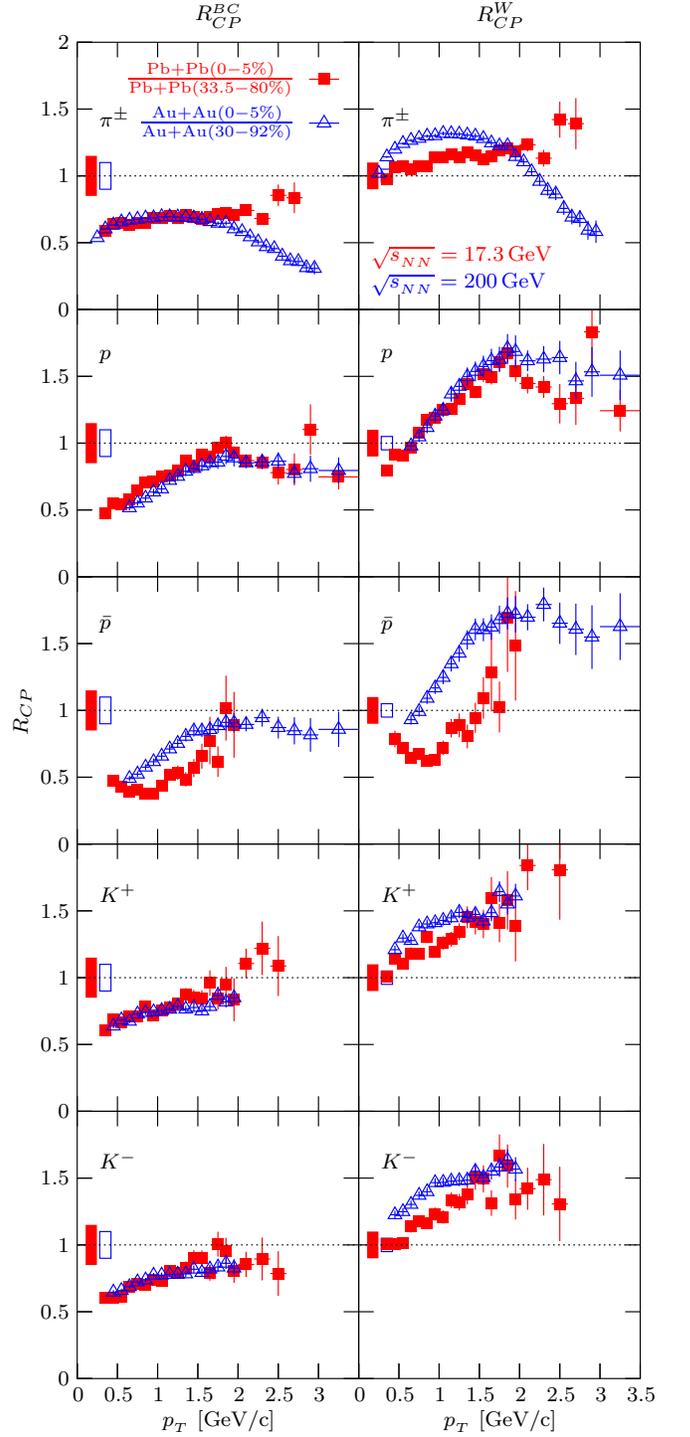

{\footnotesize\blopeps[width=8.5cm, height=19.25cm]{fig/r/Rbw5.beps}}
\caption{(Color online) Ratio of yields in central to peripheral Pb+Pb and Au+Au
collisions at $\sqrt{s_{{}_{NN}}}=17.3\GeV$ (squares) and $200\GeV$ \cite{phenixauaupid} (triangles).
Left column shows ratio $R^{BC}_{CP}$ using binary collision scaling, 
right column shows ratio $R^{W}_{CP}$ using wounded nucleon scaling. 
Centrality intervals are specified in the legend of the top left panel.
(The error bars attached to the constant $1$ line indicate normalization uncertainty.)}
\label{r2}
\end{figure}

A widely believed interpretation of the Cronin effect is momentum 
transfer from the longitudinal degrees of freedom to the transverse 
degrees of freedom by multiple scattering of partons or hadrons, depending 
on the picture. However, then this effect must 
also be present in A+A collisions, and may mask other nuclear effects. 
On the other hand the nuclear modification factor 
$R_{CP}$ would not contain most of the contribution of the Cronin effect. 
The nuclear modification factor 
$R_{CP}$ is shown in Fig.\ \ref{r2} 
for different particle types at top SPS and RHIC collision energies. 
When assuming binary collision scaling, the $\pi^{\pm}$ ratios show an 
amazing similarity at the two energies at $p_{{}_T}\leq2\GeVpc$, 
above which the $17.3\GeV$ ratio seems to stay constant (below unity), and the $200\GeV$ 
ratio begins to show a large suppression. The $p$ ratios seem to 
saturate at $p_{{}_T}\geq2\GeVpc$ at both energies and are almost identical 
over the whole $p_{{}_T}$ range. The $\bar{p}$ ratios differ 
at the two energies.
The $K^{+},K^{-}$ results are very similar at the two energies in the common $p_{{}_T}$ range. 
In fact, the NA49 $K^{\pm}$ results cover a slightly larger $p_{{}_T}$ interval than 
the published PHENIX measurements.

When using wounded nucleon scaling, 
the modification factors start from one, and show a rapid increase. 
The $\pi^{\pm},K^{\pm}$ modification factors show differences between the top SPS and RHIC 
collision energies. The difference of $\bar{p}$ results is even larger. The $p$ modification 
factors at top SPS and RHIC energies, however, are quite similar.

\begin{figure}[!ht]
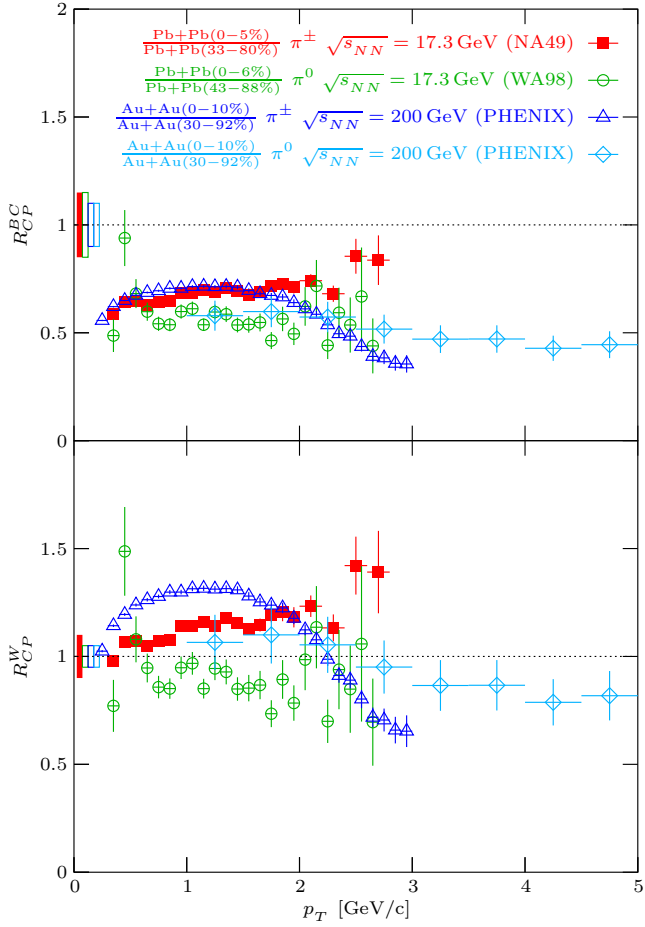

{\scriptsize\blopeps[width=8.5cm, height=12.5cm]{fig/r/Rbw7.beps}}
\caption{(Color online) Comparison of nuclear modification factors for 
$\pi^{\pm}$ and $\pi^{0}$ using
binary collision scaling ($R^{BC}_{CP}$, top) and wounded nucleon
scaling ($R^{W}_{CP}$, bottom). Results at the SPS ($\sqrt{s_{{}_{NN}}}=17.3\GeV$) in
Pb+Pb collisions from NA49 on $\pi^{\pm}$ are shown by squares, from WA98 on $\pi^{0}$
\cite{wa98} by circles. Measurements from RHIC ($\sqrt{s_{{}_{NN}}}=200\GeV$) in Au+Au
collisions for $\pi^{\pm}$ \cite{phenixauaupid} are plotted as triangles, for $\pi^{0}$ \cite{phenixpi0} as
diamonds. (The error bars attached to the
constant $1$ line indicate normalization uncertainty.)}
\label{r3}
\end{figure}

In Fig.\ \ref{r3}, the available $R_{CP}$ measurements for $\pi^{\pm}$ and $\pi^{0}$ at SPS and RHIC 
are compared \cite{wa98,phenixauaupid,phenixpi0}.
It is seen that the agreement between the result for $\pi^{\pm}$ from NA49 and 
$\pi^{0}$ from WA98 for $p_{{}_T}>0.8\GeVpc$ is quite good within errors. The 
difference between the PHENIX $\pi^{\pm}$ and $\pi^{0}$ measurements is larger 
and outside errors.

In conclusion, $R_{CP}$ ratios for identified charged particles scaled by the 
number of binary collisions are remarkably similar at SPS and RHIC energies. 
Our present results at the SPS only reach the medium $p_{{}_T}$ region, where 
soft production processes and the quark coalescence mechanism \cite{coal} 
may still dominate. It remains to be tested by future experiments, having access 
to an extended $p_{{}_T}$ range, whether energy loss of hard scattered partons 
in dense matter is also a significant process at SPS energies.


\section{Summary}

Invariant yields of $\pi^{\pm}$, $p,\bar{p}$, $K^{\pm}$ particles were 
measured in Pb+Pb
reactions at $\sqrt{s_{{}_{NN}}}=17.3\GeV$ 
collision energy around midrapidity as a function of transverse momentum up to $4.5\GeVpc$, 
with overall systematic errors below $\approx5\%$. 
Using these spectra and the previously published $\pi^{\pm}$ yields in p+p collisions,
the nuclear modification factor $R_{AA}$ for $\pi^{\pm}$ 
and $R_{CP}$ for $\pi^{\pm}$, $p$, $\bar{p}$, $K^{\pm}$ were studied.

The extracted $R_{AA}$
ratio, scaled with the number of binary collisions, shows a rapid rise with 
transverse momentum in the covered $p_{{}_T}$ region and does not show the 
strong suppression observed at RHIC.
Interestingly, the $R_{CP}$ ratios for $\pi^{\pm}, K^{\pm}$, and $p$ in the 
available $p_{{}_T}$ region stay rather close to the $\sqrt{s_{{}_{NN}}}=200\GeV$ 
RHIC results. In fact, $R_{CP}$ ratios for $p$ are almost identical.
However, although the $\pi^{\pm}$ ratio stays below one also at the SPS,
it shows a much larger suppression at RHIC above $2\GeVpc$ transverse momentum. 
When scaled with the number of wounded nucleons, the modification factors $R_{AA}$ 
and $R_{CP}$ start from one and show 
a rapid rise with $p_{{}_T}$, furthermore the $R_{pA}$ ratio stays below 
$R_{AA}$ in the covered $p_{{}_T}$ region for both SPS and RHIC collision 
energies.

The $p_{{}_T}$ range of our modification factor measurements is limited 
by the statistics of the p+p and peripheral Pb+Pb collision reference spectra.
A recently started second generation experiment, CERN-NA61 \cite{loi, prop, add}, 
will make it possible to extract reference spectra up to higher 
$p_{{}_T}$ of $4\GeVpc$ in p+p and p+Pb collisions matching the $p_{{}_T}$ 
range of the currently available central Pb+Pb data. These future measurements 
will allow a more sensitive test for the presence of high $p_{{}_T}$ particle 
suppression at SPS energy.


\begin{acknowledgements}

This work was supported by the US Department of Energy
Grant DE-FG03-97ER41020/A000,
the Bundesministerium fur Bildung und Forschung, Germany, 
the Virtual Institute VI-146 of Helmholtz Gemeinschaft, Germany,
the Polish State Committee for Scientific Research (1 P03B 006 30, 1 P03B 097 29, 1 PO3B 121 29, 1 P03B 127 30),
the Hungarian Scientific Research Fund (OTKA 68506),
the Polish-German Foundation, the Korea Science \& Engineering Foundation (R01-2005-000-10334-0),
the Bulgarian National Science Fund (Ph-09/05) and the Croatian Ministry of Science, Education and Sport (Project 098-0982887-2878).

\end{acknowledgements}



\begin{thebibliography}{99}

\bibitem{phenixauaunopid} S.~S.~Adler et al. (the PHENIX Coll.), {\it Phys. Rev.} {\bf C69} (2004) 034909.

\bibitem{phenixauaupid} S.~S.~Adler et al. (the PHENIX Coll.), {\it Phys. Rev.} {\bf C69} (2004) 034910.

\bibitem{phenixdaunopid} S.~S.~Adler et al. (the PHENIX Coll.), {\it Phys. Rev. Lett.} {\bf 91} (2003) 072303.

\bibitem{phenixdaupid} S.~S.~Adler et al. (the PHENIX Coll.), {\it Phys. Rev.} {\bf C74} (2006) 024904.

\bibitem{starauaupid} L.~Ruan (the STAR Coll.), {\it J. Phys.} {\bf G31} (2005) s1029.


\bibitem{wa98} M.~M.~Aggarwal et al. (the WA98 Coll.), {\it Eur. Phys. J.} {\bf C23} (2002) 225; 
M.~M.~Aggarwal et al. (the WA98 Coll.) Preprint [{\tt arXiv:0708.2630}]. 

\bibitem{pp} C.~Alt et al. (the NA49 Coll.), {\it Eur. Phys. J.} {\bf C45} (2006) 343.

\bibitem{enterria} D.~d'Enterria, {\it Phys. Lett.} {\bf B596} (2004) 32.

\bibitem{cronin} D.~Antreasyan et al., {\it Phys. Rev.} {\bf D19} (1979) 764.

\bibitem{na49_nim} S.~Afanasiev et al. (the NA49 Coll.), {\it Nucl. Instr. Meth.} {\bf A430} (1999) 210.

\bibitem{VCAL} C.~De Marzo et al., {\it Nucl. Instr. Meth.} {\bf 217} (1983) 405.

\bibitem{spectpaper} H.~Appelsh\"auser et al. (the NA49 Coll.), {\it Eur. Phys. J.} {\bf A2} (1998) 383.

\bibitem{VCALrel} A.~Laszlo, Time-Dependence Calibration of the Veto Calorimeter, NA49 Technical Note (2006) [{\tt EDMS:815907}]. 
NA49 Technical Notes are stored on the EDMS repository: 
[{\tt http://edms.cern.ch}].

\bibitem{venus} K.~Werner, {\it Phys. Rept.} {\bf 232} (1993) 87.

\bibitem{VCALabs} A.~Laszlo, Calculating Mean Values of Collision Parameters as a Function of Centrality, NA49 Technical Note (2007) [{\tt EDMS:885329}].

\bibitem{glissando} W.~Broniowski, M.~Rybczynski, P.~Bozek, {\it Comp. Phys. Comm.} (2007) submitted [{\tt arXiv:0710.5731}].

\bibitem{hipttech} A.~Laszlo, High Transverse Momentum Identified Charged Particle Yields in $158\GeV/\text{nucleon}$ Pb+Pb Collisions, NA49 Technical Note (2007) [{\tt EDMS:879787}].

\bibitem{gabordEdx} G.~I.~Veres, Baryon Momentum Transfer in Hadronic and Nuclear Collisions at the CERN NA49 Experiment, Ph.D. dissertation (2001) E\"otv\"os University, Budapest [{\tt EDMS:818513}].

\bibitem{dezsodEdx} D.~Varga, Study of Inclusive and Correlated Particle Production in Elementary Hadronic Interactions, Ph.D. dissertation (2003) E\"otv\"os University, Budapest [{\tt EDMS:900941}].

\bibitem{poisson} S.~Baker, R.~D.~Cousins, {\it Nucl. Instr. Meth.} {\bf A221} (1984) 437.

\bibitem{k0s} J.~B\"achler et al. (the NA49 Coll.), {\it Nucl. Phys.} {\bf A661} (1999) 45.

\bibitem{lambda} C.~Blume et al. (the NA49 Coll.), {\it J. Phys.} {\bf G34} (2007) s951.

\bibitem{compil} A compilation of particle production parameters, measured at CERN-NA49 [{\tt http://na49info.web.cern.ch/na49info/na49/
Archives/Data/NA49NumericalResults}].

\bibitem{pbpbold1} S.~Afanasiev et al. (the NA49 Coll.), {\it Phys. Rev.} {\bf C66} (2002) 054902.

\bibitem{qm05} A.~Laszlo, T.~Schuster (for the NA49 Coll.), {\it Nucl. Phys.} {\bf A774} (2006) 473.

\bibitem{sqm05} T.~Schuster, A.~Laszlo (for the NA49 Coll.), {\it J. Phys.} {\bf G32} (2006) s479.

\bibitem{qm06} A.~Laszlo, Z.~Fodor, G.~Vesztergombi (for the NA49 Coll.), {\it Int. J. Mod. Phys.} {\bf E16} (2007) 2516.

\bibitem{pbpbold2} T.~Anticic et al. (the NA49 Coll.), {\it Phys. Rev.} {\bf C69} (2004) 024902.

\bibitem{bialas} A.~Bialas, M.~Bleszynski, W.~Czyz, {\it Nucl. Phys.} {\bf B111} (1976) 461; 
A.~Bialas, W.~Czyz, {\it Acta Phys. Polon.} {\bf B36} (2005) 905.

\bibitem{phenixpi0} S.~S.~Adler et al. (the PHENIX Coll.), {\it Phys.~Rev.~Lett.}~{\bf 91} (2003) 072301.

\bibitem{coal} V.~Greco et al., {\it Phys. Rev. Lett.} {\bf 90} (2003) 202302;
R.~J.~Fries et al., {\it Phys. Rev. Lett.} {\bf 90} (2003) 202303;
R.~C.~Hwa, C.~B.~Yang, {\it Phys. Rev.} {\bf C70} (2004) 024904.

\bibitem{loi} N.~Antoniou et al. (the NA61 Coll.), Study of Hadron Production in Collisions of Protons and Nuclei at the CERN SPS, NA49-future Letter of Intent (2006) [{\tt CDS:CERN-SPSC-2006-001, SPSC-I-235}].
CERN documents are stored on the CDS repository: 
[{\tt http://cds.cern.ch}].

\bibitem{prop} N.~Antoniou et al. (the NA61 Coll.), Study of Hadron Production in Hadron-Nucleus and Nucleus-Nucleus Collisions at the CERN SPS, NA49-future Proposal (2006) [{\tt CDS:CERN-SPSC-2006-034, SPSC-P-330}].

\bibitem{add} N.~Antoniou et al. (the NA61 Coll.), Additional Information Requested in the Proposal Review Process, Addendum to the NA49-future Proposal (2007) [{\tt CDS:CERN-SPSC-2007-004, SPSC-P-330}].

\end{thebibliography}
\end{document}